\documentclass[aps,prl,twocolumn,10 pt,superscriptaddress,showpacs]{revtex4}
\usepackage{amssymb}
\usepackage{graphicx}

\begin{document}

\title{A study of flux lines lattice order and critical current with time of flight small angle neutron scattering.}
\author{Alain Pautrat}
\affiliation{Laboratoire CRISMAT, UMR 6508 du CNRS, ENSICAEN et Universit$\acute{e}$ de Caen, 6 Bd Mar$\acute{e}$chal Juin, F-14050 Caen 4, France.}
\author{Annie Brulet}
\affiliation{Laboratoire L\'{e}on Brillouin, UMR 12 du CNRS et du CEA, \\ CEA Saclay, 91191 Gif/Yvette, France.}
\author{Charles Simon}
\affiliation{Laboratoire CRISMAT, UMR 6508 du CNRS, ENSICAEN et Universit$\acute{e}$ de Caen, 6 Bd Mar$\acute{e}$chal Juin, F-14050 Caen 4, France.}
\author{Patrice Mathieu}
\affiliation{Laboratoire Pierre Aigrain de l'Ecole Normale Sup\'erieure, UMR 8551 du
CNRS, associ\'ee aux universit\'es Paris 6 et 7, 75231 Paris Cedex5, France.}

\begin{abstract}
Small angle neutron scattering (SANS) is an historical technique to study the flux lines lattice (FLL) in a superconductor.
Structural characteristics of the FLL can be revealed, providing fundamental information for the physics of vortex lattice.
 However, the spatial resolution is limited and all the correlation lengths of order are difficult to extract with precision. 
 We show here that a time of flight technique reveals the Bragg peak of the FLL, and also its translational order with a better resolution.
 We discuss the implication of these results for pinning mechanisms in a Niobium sample.
\end{abstract}

\pacs{61.05.fg,74.25.Qt,74.70.Ad}
\newpage
\maketitle

\section{Introduction}

 The existence of a flux lines lattice (FLL) in a superconductor of the second kind was initially confirmed by Cribier et al using small
 angle neutron scattering (SANS) \cite{cribier}. Afterwards, SANS was an essential technique to study the physics of the FLL,
 including FLL transitions \cite{ted,ling,morten,bi}, pinning mechanisms \cite{gammel,surface}, moving FLL \cite{thorel,echo}
 and out of equilibrium features \cite{nbse2}.
A central question for the physics of FLL is the nature of the FLL in the presence of unavoidable structural disorder.
 It is clear that the flux lines are forming planes which are
 ordered at a sufficient long range to display Bragg peaks in the diffraction pattern.
By contrast, a genuine glassy or liquid state presents a large spread of Bragg angles and a degenerated orientational
 order that lead to diffuse rings of scattering in the reciprocal space. However, it is not clear whether the FLL is ordered at very long range 
\cite{klein} or is fracturing
 at the intermediate scale \cite{gautam}. The result is important in the field
 of disordered elastic media where novel phases, not existing for the crystalline matter (i.e the Bragg Glass) \cite{giam}, have been proposed.
 It is also important for
 discriminating between pinning effects,
 since FLL order and bulk critical current can be linked in the framework of the elastic theory of collective pinning \cite{LO}. On the 
other hand, pinning at the surface
 is also very efficient but the associated critical current is not, at least not directly, related to the disorder in the bulk FLL
 structure \cite{surfpinning}.
 SANS is the sole technique from which the bulk FLL correlation lengths
 can be in principle extracted \cite{gammel}. 
\newline
%For analysing FLL structures,
% the discriminating parameter is the structure factor $S(Q)$ which is the Fourier transform of the correlation function $C(r)$.
% For a Bragg glass with quasi-long range order, the correlation function decays algebraically with the distance
%yielding $C(r) \propto r^{\eta}$, with $\eta\sim 1$. For more classical form of disorder, $C(r)\propto exp^{-r/ \ell}$ with $\ell$ the correlation length
% of positional order.
%The scattering intensity $I(Q)$ in a diffraction experiment is related to the structure factor $S(Q)$ which is the Fourier transform 
%of the correlation function $C(r)$, i.e. the quantity of interest for the positional order.
 Experiments are usually performed with the neutron beam applied along the magnetic field, the so-called longitudinal geometry.
The resolution is good enough in the longitudinal direction to extract the longitudinal correlation length (the straightness of flux lines)
 with accuracy. Note however that this length is directly affected by field lines bending due to the demagnetization field \cite{nopeak}
 or by the self field if a transport current is applied \cite{surface}. Its interpretation is then tricky \cite{nbse2}.
The scattering intensity $I(Q)$ contains the square of the modulus of the structure factor $S(Q)$,
 which is the Fourier transform 
of the positional correlation function $C(r)$. $I(Q)$ is analyzed after a radial averaging of the intensity in the detector plane.
In this direction however, the resolution function of the small
angle diffractometer is strongly dominating \cite{bob, christen} and a direct information can not be obtained from the Bragg peak shape. 
 As a consequence, indications that the FLL is in a Bragg glass state arises from the analysis of the decrease of the scattered intensity with the magnetic field \cite{klein}, and 
not from the expected power-law decay \cite{reverse}.
  For the same reasons, the broadening of the peak in the radial direction which gives informations on the crystallite size
 can not be extracted without removing the dominating contribution of the resolution function. This leads to a 
large uncertainty in the size of crystallites.
\newline
 The other geometry is the transverse geometry where the beam is perpendicular to the magnetic field, i.e. the high resolution direction
corresponds to the orientational order. A major drawback is that the Bragg planes are difficult to bring in diffraction conditions
 for any misalignment between FLL and the applied field \cite{christen}. In this geometry, some evidence
 of a power law decay of Bragg peaks are observed in a Niobium of good quality, what is consistent with a Bragg Glass phase \cite{reverse}.
 However, the intrinsic width of these Bragg peaks were much larger than the experimental resolution. 
This implies that the FLL fractures at an intermediate length scale, typically in the micrometer range.
 These results are not expected in the Bragg-Glass scenario of quasi-long
 range order where Bragg peaks have no intrinsic width in the regime of power law decay. 
Complementary experimental data should be helpful to clarify
 the situation.
 
We have used another technique to extract the FLL crystallite size, using the spectrometer
 with the usual longitudinal geometry but in the time of flight (TOF) mode.
The time of flight of neutrons between the chopper source and the detector is used to separate in the scattering
 intensity the neutrons of different wavelengths.
In the conventional steady state mode, a monochromatic beam can be selected only with a large wavelength
 spread $\Delta \lambda / \lambda \sim 10-20$ \% which has an important contribution in the resolution function.
 In the TOF mode,
 this wavelength spread arising from the pulse and the channel widths can be negligible. 
Another advantage of the TOF is that scattering to higher angles $\theta$ which gives a
better resolution \cite{CD} is direct thanks to the large range of wavelengths available in the single experiment. The
 drawback is that the neutron flux is low at large wavelength due its the Maxwellian spectrum (the wavelength
 at maximum flux is here $\lambda\approx 2$ \AA). More practical but also interesting, it is not necessary to rock
 the sample through the Bragg conditions which are satisfied at a fixed position when $\lambda$ is continuously tuned 
($\theta_{Bragg}=Q_{Bragg}\lambda/2\pi$ with $\theta_{Bragg}$ the scattering angle).
 The definition of a Bragg peak is finally largely improved, i.e. more data points are available to describe it what
 improves significantly the accuracy on the peak width.

SANS data were measured with the PAXY small-angle diffractometer (Laboratoire Leon Brillouin, Saclay, France), which can be used in both steady state and TOF modes.
The sample under study is a slab of pure Niobium with critical temperature $T_c$=9.2 K, Ginzburg-Landau parameter $\kappa\approx 1$ 
and second critical field $B_{c2}$=0.385 T at T=2 K. No peak effect in the critical current was observed, what is consistent with a 
sample of the best homogeneity \cite{nopeak}.

Before the TOF measurements, the FLL was measured with the conventional steady state set-up, using a mechanical velocity
 selector which chooses the neutrons wavelength $\lambda$ with a Gaussian distribution
 of full width at half maximum (FWHM) $\Delta \lambda$= 0.11 $\times$ $\lambda$
 (or a variance $\sigma_{\lambda}^2$ given by $\Delta \lambda/2\sqrt{2ln2}$= $\sigma_{\lambda}$).
The wavelength was fixed at $\lambda$=10 $\AA$.  
$I(Q)$ is obtained after regrouping the different rocking curves and after a radial averaging of the intensity.
We observe in fig.1 the Bragg peak associated with the FLL and centered at $Q_{FLL}=2\pi / d_{FLL}$
 ($d_{FLL}= 1.155 a_0$, $a_0=1.074 \sqrt{\phi_0/B}$ is the unit cell of the hexagonal lattice).
The Bragg peak is well fitted by a Gaussian because dominated by the spectrometer resolution \cite{pedersen}.
This latter is related to the angular distribution of the incident neutron beam at the detector position,
 to the wavelength distribution $\Delta \lambda$, and the detector resolution $\Delta $R.
Adding the different Gaussian contributions to the resolution gives:
 
\begin{equation}
\sigma^2_{Q_{res}} = (2\pi/\lambda)^2 \sigma_{\theta}^2+(Q/\lambda)^2 \Delta \lambda^2/(2\sqrt{2ln2})^2.
\end{equation}

The detector resolution contribution being negligible compared to the one arising from the incident neutron beam,
 a good approximation of the angular variance $\sigma_{\theta}$ is \cite{seeger}
 
\begin{equation}
\sigma_{\theta}^2 \approx \sigma^2_{\theta_{beam}} \approx 1/4 (D_1/2.L_1)^2 + 1/4 [D_2.(L_1+L_2)/2.L_1.L_2]^2
\end{equation}

with $D_1$, $D_2$ the aperture diameters, $L_1$ the distance between the two apertures and $L_2$ between $D_2$ and the detector
 (Here D1=12 mm, D2=7 mm, L1=4750 mm, L2=6870 mm).

With our set-up, $\sigma_{Q_{res}}\approx 6.8$ $10^{-4}$ $\AA^{-1}$ at $Q= 6.64$ $10^{-3}$ $\AA^{-1}$. To measure the
 intrinsic width of the Bragg peak, it is necessary to 
deconvolve the experimental data $\sigma_{Q_{exp}}$ from $\sigma_{Q_{res}}$ \cite{bob,seeger}.
 As shown in fig.1, $\sigma_{Q_{exp}}\approx \sigma_{Q_{res}}$ and we can only conclude that the FLL crystallite size has a minimum value of some $\mu$m.
 We have also performed full rocking curves of Bragg peaks whose FWHM are related to the perfection of flux lines along the field
 direction. We measure resolution limited widths for B= 2000, 2500, 3000 and 3500 G,
 indicating a longitudinal correlation length of more than 100 $\mu$m, i.e. straight flux lines. Orientational order was not changing in the field range investigated. 
 
SANS experiments in the TOF mode were then performed. The velocity selector was removed and a chopper with one slit (12 $\times$ 1 mm$^2$)
 was placed before the cryomagnet containing the sample.
 The chopper to detector distance was $D_{chopper}$=7110 mm.
 The collimation and the sample to multidetector distance were kept. The total pulse width $\tau_{chopper}$=450 $\mu$s is given
 by the rotation of the 1 mm
 width slit of the chopper in front of the 7 mm diameter sample diaphragm. The flying 
times of neutrons were analyzed in times frame of 256 TOF channels of $\tau$=150 $\mu$s each.
 The scattering intensity of the sample was recorded at T=2K for different magnetic fields in the superconducting state (field cooling procedure) and
 a background was measured in the normal state (B=4000 G$>$B$_{c2}$). In the TOF mode, the neutrons are recorded as function of the time of
flight $t$ for different angles $\theta$, then $t$ is converted to effective wavelengths, giving I($\theta$, $\lambda$).
 Each set of raw scattering data was corrected for the detector efficiency, the sample transmission and the wavelength distribution of the incident beam flux by dividing each scattering data by I($\theta$=0, $\lambda$). Data obtained in the normal state was used as the background scattering for data obtained in the superconducting state.
 Typical data showing the FLL Bragg peak as function of ($\theta$,$\lambda$) are shown in fig.2.
 In TOF mode, the contribution of $\Delta \lambda$ in equation (2) is no more coming from the
 wavelength distribution delivered by a velocity selector but arises from the short pulse and the TOF channel width.
Using $\lambda=h/(m_nv)$ ($h$ is the Planck constant, $m_n$ the neutron mass and $v$ the neutron velocity) gives
 $\Delta \lambda_{chopper}=\tau_{chopper}/(0.253 \times D_{chopper})$ and $\Delta \lambda_{\tau}=\tau /(0.253 \times D_{chopper})$
  Here, $\Delta \lambda_{chopper}$ and $\Delta \lambda_{\tau}$ are in \AA, $\tau_{chopper}$
 and $\tau$ in $\mu$s, $D_{chopper}$ in mm.
Assuming Gaussian distributions for these two contributions, the term $(\Delta \lambda/\lambda)^2$ in equation (1)
 becomes $(\Delta \lambda_{chopper}/\lambda)^2+(\Delta \lambda_{\tau}/\lambda)^2$.
It is much smaller and negligible compared to the contribution arising from the angular divergence of the beam. 
  The position of the peak changes with ($\theta$, $\lambda$) according to the Bragg law as shown in the inset of fig.2.
The width of the peak decreases at large $\theta$, as expected from equation (1).
 In the TOF mode, interestingly, this $\theta$ dependence can be analyzed in a single measurement when measuring I($\theta$, $\lambda$).
\newline 
A typical size of crystallites can be estimated using the Scherrer formula \cite{scherrer}:
\begin{equation}
 \Delta \theta_{crystal} \approx K.\lambda/S.cos\theta
\end{equation}
with $K$ the Scherrer constant of the order of unity \cite{notebene2} and $S$ the mean size of the crystallite. 
Note that $S$ is an effective length, measured in the direction of the diffraction vector.
 
Finally, the FLL Bragg peak broadening varies as:
\begin{equation}
\sigma_{Q_{FLL}}^2\approx (\pi.\sigma_{\theta_{beam}})^2/(d_{FLL}.\theta)^2+(2.\pi.K )^2/(cos\theta.S)^2
 \end{equation}
 Since $cos \theta \approx 1$ at small angle, it can be rewritten in the compact form
 \begin{equation}
\sigma_{Q_{FLL}}^2 \approx A/\theta^2 +B/ S^2
 \end{equation}
 with $B$ a constant of the order of $(2\pi)^2$, and $A$ is function of the Bragg planes spacing.

Despite the gain in resolution offered by the TOF, the intrinsic width of the peak
 still corresponds to a low contribution compared to the instrumental resolution.
The equation (5) has in principle two adjustable parameters, $A$ and $B/S^2$ (the instrumental resolution due to the intercept of the incident beam
 in the detector plane and the crystallite size $S$). Since the prefactor $A$ in the beam resolution contains only geometrical parameters,
 it can be calculated and then fixed at the expected value. In such a case, the fit has only one free parameter.
 In fig.3, we show the comparison between the two fitting procedures, with one or two adjustable parameters
The fits are of equivalent quality. From the fitting curves at different fields (fig.4), we can deduce $S$ as function of the magnetic field values.
 As shown in fig. 5, the order of magnitude of $S$ and its
 field dependence are  found similar for the two procedures, albeit with larger error bars when using the two
 parameters fit. It certainly reflects the relative uncertainty of the extraction of $S$,
 but shows also that the order of magnitude is correct. In the following, we will discuss the case of $S$ deduced from the one parameter fit,
 but the results and discussion are similar with either fitting procedures.

We observe that $S$ increases with the magnetic field, meaning that the FLL becomes more perfect as its density increases. 
$S$ is however slightly larger at 500 G than at 1000 G.
 In a low kappa superconductor such as Niobium, the first critical field is large, i.e $B_{c1}\sim 1500 G$ at 2K. The intermediate
 mixed state
 where FLL and Meissner regions coexist is extended at low field, specially with the slab geometry of the sample with a
 large demagnetizing factor \cite{brandt}. The measurements at the lowest field are then likely influenced
 by the increasing presence of Meissner domains in the sample.

In the mixed state, the critical current density $J_c (bulk)$ is related
 to the positional correlation length $R_c$ in the Larkin Ovchinikov (LO) model \cite{LO}, which is valid for short lenght scales displacements,
 typically $r_p$, the one of the pinning potential \cite{lark}. 
It has been pointed out that a diffraction experiment probes scale of the order of the lattice spacing \cite{comment}.
 The correlation length $S$ can not be directly compared with $R_c$, but should be rescaled as $S \approx R_c (d_{FLL}/r_p)^2$ \cite{comment}.
$r_p$ is expected to vary between the flux line core size $\xi$ for isolated flux lines up to some fraction of the lattice spacing
 $d_{FLL}$ when flux lines largely overlap \cite{brandtpinning}.
We have calculated $J_c(bulk)\approx C_{66} r_p/(4 B R_c^2)$  \cite{LO} with the two limiting values of $r_p$.
The shear modulus $C_{66}$ formula was given in \cite{mumut}, and $\xi (2K) \approx (\phi_0/2\pi B_{c2})^{1/2} \approx$ 29 nm.
$J_c(bulk)$ is finally reported in Fig.6, for $r_p=\xi$ and for $r_p=d_{FLL}$.
\newline 
The next step is to compare with the real critical current $J_c(exp)$ of the sample.
It was measured on a small piece of the same Nb sample (width $\times$ length $\times$ 
thickness = 0.1 $\times$ 0.3 $\times$ 0.02 $cm^3$), using the irreversible part of the magnetization
and applying the Bean model for a slab geometry. The resulting $J_c(exp)$ is shown in Fig.6. Clearly, $J_c(bulk)$ is very different
 than $J_c(exp)$ for $B/B_{c2} >$ 0.4, in the regime where 
our Niobium sample is clearly in the pure mixed state, whatever the $r_p$ value.

We conclude that a large amount of critical current is not coming from a bulk pinning contribution of the LO type.
Another possibilty is that we are measuring a crystallite size unrelated to any bulk pinning mechanism, for reasons which are not clear.
In soft superconductors of the second kind, the bulk contribution to the pinning can be very low and most of the critical current
 can arise from a surface origin. In particular, surface pinning is known to be a realistic source of pinning in Niobium \cite{surfpinning, surf}.
 This pinning mechanism is based on equilibrium equations and boundary conditions for flux lines
 over a realistic (rough) surface \cite{MS}. The surface contribution to the pinning gives a critical current $i_c (A/m)= \varepsilon. sin \theta$ 
where $\varepsilon$ is the vortex potential
 (i.e. the reversible magnetization) and $\theta$ is a critical angle characterizing the surface roughness. $i_c$ is a superficial current
 but can be rewritten
 as a critical current density $J_c(surf)(A/m^2)=2 i_c/t$ where $t$ is the sample thickness and the factor 2 stands for the two surfaces
 perpendicular to the magnetic field.
standard values of $\theta$ are around few degrees. To estimate $i_c$, $\varepsilon$ was computed as function of $B$ using numerical calculation
 following \cite{brandt2}.
For this calculation, we have used a Ginzburg Landau parameter $\kappa=1$ and $B_{c2}=0.385$ T as observed experimentally.
 The result is reported in Fig.7.
 A direct observation is that the magnetic field dependence of $\varepsilon$ and $J_c(exp)$ are quite similar, as expected 
if the critical angle is not much depending on the value of the field.
Fitting the experimental $J_c(exp)$ of our Niobium sample
 with the surface pinning expression $J_c(surf)= 2/t \times \varepsilon \times sin\theta$ gives a very good agreement
 with a critical angle $\theta=4$ deg (Fig.8). This is 
a large, but still reasonable value for a bulk sample with unpolished surfaces \cite{nous}. We do not exclude that edge currents
 play also a role for the critical current. Finally, the good agreement between the surface pinning model 
and the experimental data, in addition to the neutron scattering experiment, allow to conclude that surface critical currents
 are likely dominant in this sample.  
 \newline
Another important result of the TOF experiment is that the FLL is fracturing in the micrometer scale, as found using reverse monte carlo
 analysis in a different geometry \cite{reverse}. This result is \textit{a priori} not expected in the framework of purely elastic models
 where quasi long range ordering is expected in clean samples such as Niobium. Note that we have used field cooled, zero field cooled and
 the so-called shaking procedures to
 induce a better positional order of the FLL \cite{andrew}. The widths of the peak are not changed (within resolution),
 meaning that unpaired dislocations are likely not responsible for the Bragg peak width and that the FLL is close to equilibrium. 

To conclude, we have measured the crystallite size of the FLL in Niobium using neutron scattering with a time of flight mode.
Crystallite sizes are in the micrometer range, and increase with the field. These results show that the FLL positional order gets better when the
 flux line density increases. The crystallite sizes are found unrelated to the critical current using a bulk collective pinning approach. This implies other source of pinning unrelated to the bulk FLL structure,
 likely surface pinning which is found to describe quantitatively the critical current values.

\begin{figure}[t!]
\begin{center}
\includegraphics*[width=8.0cm]{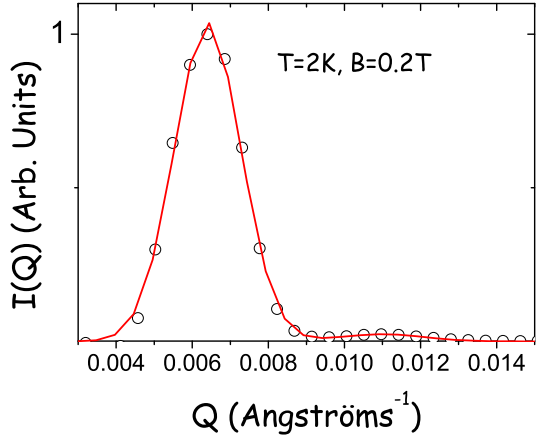}
\end{center}
\caption{Normalized intensity as function of the scattering vector $Q$ ($\lambda=10 \AA$, T=2K)).
 The first Bragg peak is fitted by a Gaussian of center $Q$= 0.0066 $\AA^{-1}$ and FWHM 0.00165 $\pm$ 0.00004 $\AA^{-1}$ 
(solid line), very close to the calculated FWHM due to the resolution (0.00160 $\AA^{-1}$).}
\label{fig.1}
\end{figure}

\begin{figure}[t!]
\begin{center}
\includegraphics*[width=8.0cm]{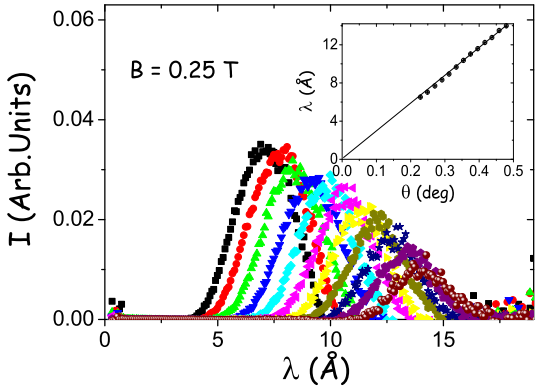}
\end{center}
\caption{Intensity of the FLL Bragg peak $I$ as function of $\lambda_{neutrons}$ (time of flight mode). Each Bragg peak
 corresponds to a different $\theta$
($\theta$=0.201, 0.229, 0.249, 0.271, 0.291, 0.312, 0.333, 0.354, 0.375, 0.396, 0.412, 0.437 deg). In the inset is shown $\lambda_{peak}$,
 the center of the Bragg peak, as function of the angle $\theta$. The solid line is the Bragg law,
with $Q_{Bragg}$= 7.47 10$^{-3}$ $\pm$ 0.02 \AA $^{-1}$ (this is the value expected for the hexagonal lattice at 2500 G within
 the resolution of the magnet).}
\label{fig.2}
\end{figure}

\begin{figure}[t!]
\begin{center}
\includegraphics*[width=8.0cm]{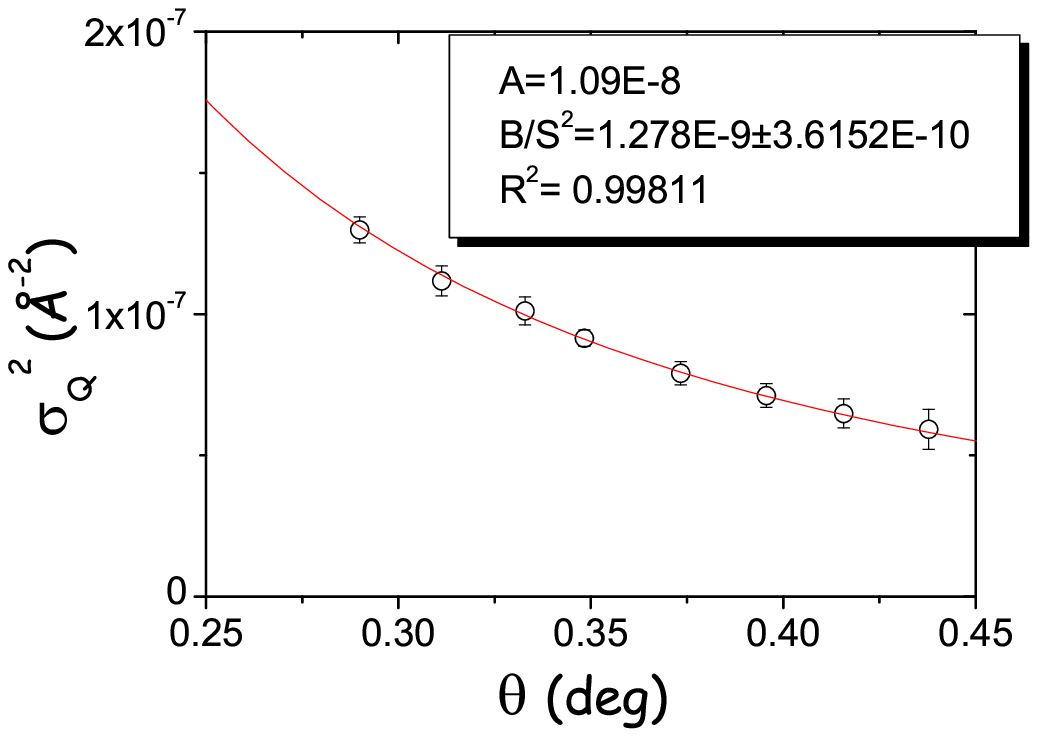}
\end{center}
\begin{center}
\includegraphics*[width=8.0cm]{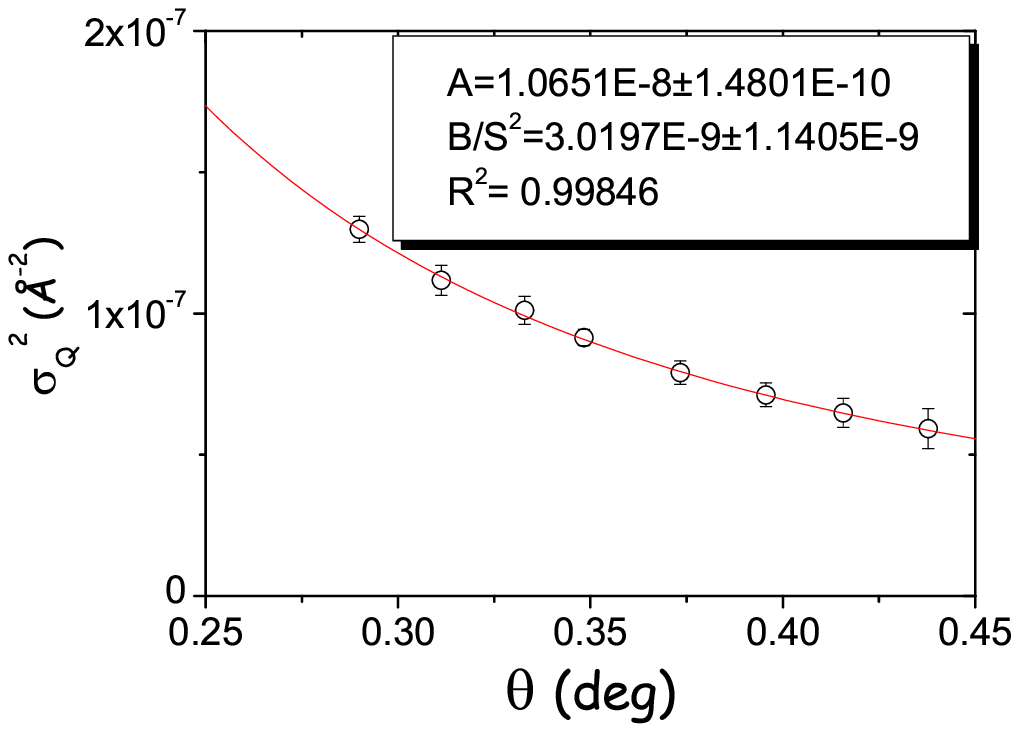}
\end{center}
\caption{$\sigma^2_{Q_FLL}$ as function of the angle of scattering $\theta$ (T=2K, B=2000 G).
 In both graphs, the solid line is a fit with the equation $\sigma^2_{ Q_{FLL}} \approx A/\theta^2 +B/ S^2$,
 with one free parameter $B/S^2$ (left) or with two free parameters $A$ and $B/S^2$ (right) (see text). $R^2$ is the coefficient of determination.}
\label{fig.3}
\end{figure}

\begin{figure}[t!]
\begin{center}
\includegraphics*[width=8.0cm]{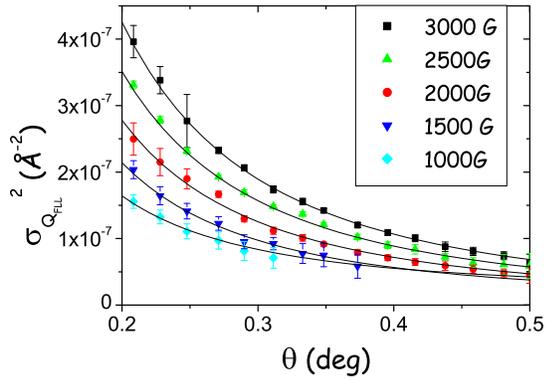}
\end{center}
\caption{$\sigma^2_{Q_FLL}$ as function of the angle of scattering $\theta$ (T=2K, different field values).
 The solid line is a fit with the equation $\sigma^2_{ Q_{FLL}} \approx A/\theta^2 +B/ S^2$, with $B/S^2$ as a free parameter(see text). For clarity, the fit with two parameters is not shown.}
\label{fig.4}
\end{figure}

\begin{figure}[t!]
\begin{center}
\includegraphics*[width=8.0cm]{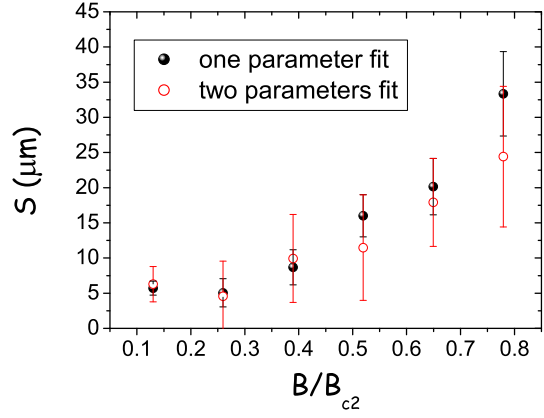}
\end{center}
\caption{Size of the vortex crystallite as function of the reduced magnetic field $B/B_{c2}$ (T=2K, B$_{c2}$=3850 G)
 obtained with the single parameter fit or with the two parameters fit. }
\label{fig.5} 
\end{figure}

\begin{figure}[t!]
\begin{center}
\includegraphics*[width=8.0cm]{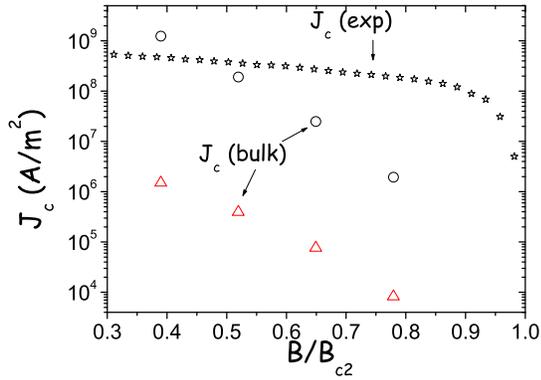}
\end{center}
\caption{Experimental critical current measured at T=2K (stars), compared with
 the critical current calculated with the LO model and $S$ values as explained in the text (empty points correspond to $r_p=\xi$ and empty triangles to $r_p=d_{FLL}$).}
\label{fig.6}
\end{figure}

\begin{figure}[t!]
\begin{center}
\includegraphics*[width=8.0cm]{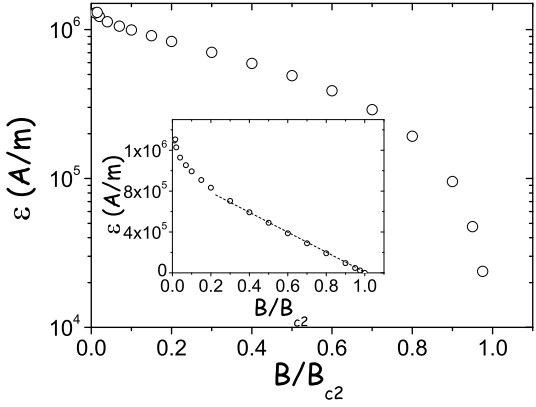}
\end{center}
\caption{Potential $\varepsilon$ (or reversible magnetization) computed using the Brandt approach \cite{brandt2} as function of reduced magnetic field in a semi
-log scale ($\kappa$=1 and $B_{c2}=$ 0.385 T).
 In the inset is shown the same graph in a linear scale. The dotted line is the Abrikosov line.}
\label{fig.7}
\end{figure}

\begin{figure}[t!]
\begin{center}
\includegraphics*[width=8.0cm]{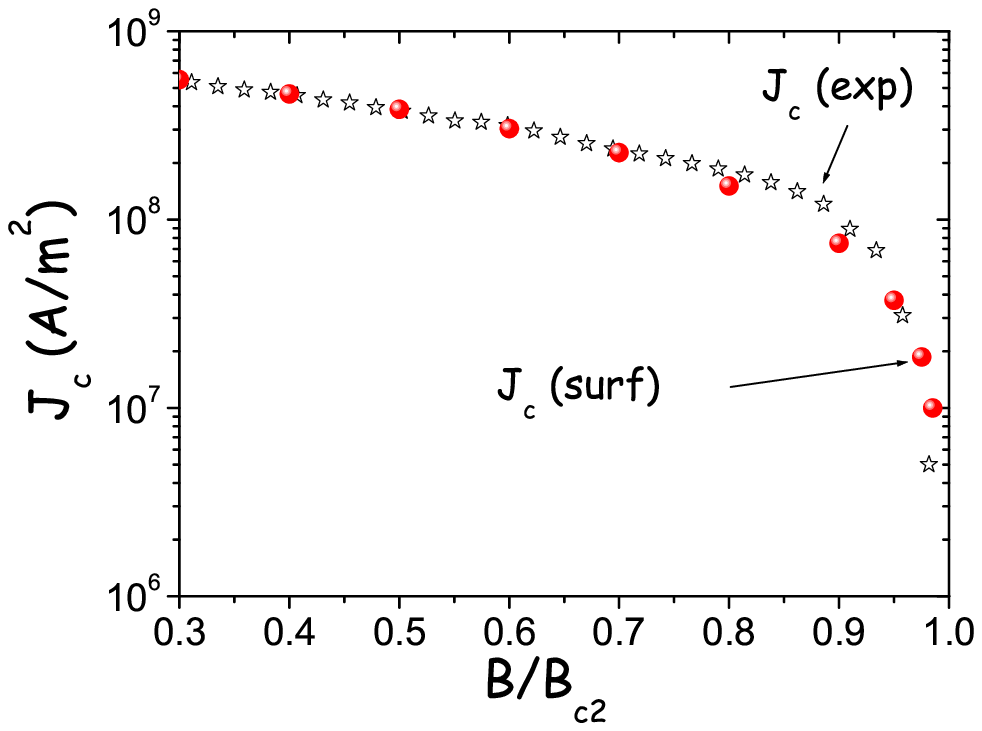}
\end{center}
\caption{Experimental critical current measured at T=2K (stars), compared with
 the critical current calculated with the surface pinning model with a critical angle $\theta=4$ deg (plain points).}
\label{fig.8}
\end{figure}

\begin{references}
\bibitem{notebene2} The Scherrer constant $K$ is a function of the crystallite shapes but corrections are not important
 for the present discussion ($K$ is between 0.6 and 2), see \cite{scherrer}.
\end{references}
\end{document}